\documentclass{article}
\usepackage{pgfplots}
\usepackage{lipsum}
\usepackage{caption}
\usepackage{subcaption}
\usepackage{authblk}
\usepackage[english]{babel}
\usepackage{amsfonts}
\usepackage{amsmath}
\usepackage[top=2cm, bottom=2cm, left=2cm, right=2cm]{geometry}
\usepackage{fancyhdr}
\usepackage{multicol}
\usepackage{bbold}
\usepackage{epsfig}
\usepackage{graphics}
\usepackage{epsfig}
\usepackage{lipsum}
\usepackage{fancyhdr}
\usepackage{booktabs}
\usepackage{cleveref}
\usepackage{wrapfig}
\usepackage[export]{adjustbox}
\usepackage{rotating}
\usepackage{amsmath}

\usepackage{cite}

\renewenvironment{abstract}{
	
	\hfill\begin{minipage}{0.95\textwidth}
		\rule{\textwidth}{1pt}}
	{\par\noindent\rule{\textwidth}{1pt}\end{minipage}}


\begin{document}

	\title{\textbf{Bipartite and tripartite entanglement in an optomechanical ring cavity}}
	\author[1]{\textbf{Oumayma El Bir}}
	\author[2,3]{\textbf{Abderrahim Lakhfif}}
	\author[2,3]{\textbf{Abdallah Slaoui}}
	\affil[1]{\small ESMaR, Faculty of Sciences, Mohammed V University in Rabat, Rabat, Morocco.}
	\affil[2]{\small LPHE-Modeling and Simulation, Faculty of Sciences, Mohammed V University in Rabat, Rabat, Morocco.}
	\affil[3]{\small Centre of Physics and Mathematics, CPM, Faculty of Sciences, Mohammed V University in Rabat, Rabat, Morocco.}

	\maketitle
	\begin{center}
		\textbf{Abstract}
	\end{center}
	\begin{abstract}

Entanglement serves as a core resource for quantum information technologies, including applications in quantum cryptography, quantum metrology, and quantum communication. In this study, we give a unifying description of the stationary bipartite and tripartite entanglement in a coupled optomechanical ring cavity comprising photon and phonon modes. We numerically analyze the stationary entanglement between the optical mode and each mechanical mode, as well as between the two mechanical modes, using the logarithmic negativity. Our results demonstrate that mechanical entanglement between the two mechanical modes is highly dependent on the optical normalized detuning and the mechanical coupling strength, with entanglement maximized within specific detuning intervals and increased coupling broadening the effective range. Furthermore, we study the entanglement's sensitivity to temperature, noting that higher coupling strengths can sustain entanglement at elevated temperatures. The study also reveals that the entanglement between the mechanical mode and the optical mode is enhanced with increasing laser power, but is similarly susceptible to thermal noise. Additionally, we explore tripartite entanglement through the minimum residual contangle, highlighting its dependence on detuning, temperature, and laser power. Our findings underscore the importance of parameter control in optimizing entanglement for quantum information processing applications.

	\end{abstract}
	
	\vspace{0.5cm}
	\textbf{Keywords}: quantum entanglement; optomechanics; ring cavity; decoherence; tripartite entanglement
	
	\section{Introduction}

\indent Quantum entanglement is a fundamental phenomenon in quantum mechanics where quantum systems can be correlated in such a way that the state of one system instantly influences the other one, no matter the distance between them. Initially highlighted by Einstein, Podolsky, and Rosen (EPR) in 1935 \cite{einstein1935can}, entanglement as a non-local correlation challenges classical notions of locality and causality. Experimental demonstrations through Bell's inequalities have reinforced its foundational role in quantum mechanics \cite{bell1964on,aspect1981experimental,aspect1982experimental,aspect1982experimental2}. Today, entanglement is recognized as a vital resource for various quantum technologies applications, including quantum teleportation \cite{bennett1993teleporting,Dakir2023,Kirdi2023}, superdense coding \cite{bennett1992communication}, quantum cryptography \cite{ekert1991quantum}, quantum computing and communication \cite{pathak2013elements,Slaoui2024,ACzerwinski2022}, and quantum networks \cite{kimble2008quantum}.\par

Optomechanical systems, which involve the interaction of electromagnetic field with mechanical motion, represent a significant frontier in merging quantum information science with macroscopic mechanical systems \cite{aspelmeyer2014cavity,LakhfifSlaoui2024}. In these systems, the interaction between light and mechanical oscillators is achieved through radiation pressure. This interaction can be harnessed to generate entanglement between optical and mechanical modes, providing a platform for exploring quantum phenomena in systems with substantial mass and size \cite{palomaki2013entangling,Ockeloen-Korppi2018,Riedinger2018}. The ability to entangle mechanical oscillators with optical fields has profound implications for quantum sensing and precision measurement, as entangled states can surpass classical measurement limits \cite{Li2021cavity,barzanjeh2022optomechanics,Czerwinski2022,barzanjeh2019stationary}. Furthermore, entangling multiple mechanical oscillators could enable the development of hybrid quantum networks that integrate mechanical components, thus enhancing the versatility of quantum technologies \cite{rakhubovsky2019entanglement}. These developments are significant for both fundamental quantum mechanics research and practical applications in emerging quantum technologies.

Cavity optomechanics has garnered considerable focus in recent years due to its potential in precision measurement \cite{xu2022optomechanically,li2021cavity,brawley2016nonlinear}, quantum information processing \cite{kippenberg2007cavity,sohail2020enhancement,marquardt2011quantum,el2023mirrors,el2020quantum,Lakhfif2022pairewise,Lakhfif2022dynamics}, and fundamental tests of quantum mechanics \cite{chen2013macroscopic,aspelmeyer2012quantum,marquardt2009optomechanics,milburn2011introduction}. Among the various configurations, ring optomechanical cavities, characterized by their ring-shaped optical resonators coupled with mechanical mirrors, present a promising platform for exploring the interaction between light and mechanical motion \cite{huang2009entangling,schulze2010optomechanical}. These systems have shown exceptional potential due to their unique topology, which enhances the coupling efficiency and interaction strength between optical and mechanical modes \cite{huang2009entangling}. This enhancement can lead to more robust and easily controllable entangled states. A ring optomechanical cavity typically consists of a ring-shaped optical resonator with a movable mechanical oscillator that interacts with the electromagnetic field closed in the cavity through the radiation pressure force. The momentum transfer from photons to the mechanical phonons causes a small displacement of the mechanical oscillator, which in turn changes the length of the cavity and modifies its state. This interaction leads to phenomena such as optomechanical cooling, electromagnetically induced transparency, and the generation of non-classical states of light \cite{kippenberg2008cavity,Elsasser2003collective,Ma2013electromagnetically,schliesser2009resolved}. In this context, we aim to shed light on one of the most significant and intriguing aspects of ring optomechanical cavities, which is their potential to manipulate and generate entangled states. Entanglement is a quintessential feature for many applications in quantum information processing and quantum computing \cite{sarma2021continuous,tang2022perspective,gut2020stationary,chen2017entanglement}. In ring optomechanical systems, entanglement can be achieved between optical and mechanical modes, or even between multiple mechanical oscillators. In what follows, we provide a comprehensive analysis of the dynamics of a ring optomechanical cavity. We start by describing the model, presenting the Hamiltonian, and then determine the system dynamics by solving the differential equations of motion, which leads to the covariance matrix of the optomechanical system. Subsequently, we will discuss the behavior of quantum entanglement present in the system using logarithmic negativity to measure bipartite entanglement and minimum residual contangle to quantify tripartite entanglement. Finally, we will end the paper with a few concluding remarks.
%%%%%%%%%%%%%%%%%%%%%%%%%%%%%%%%%%%%%%%%%%%%%%%%%%%%%%%%%%%%%%%%%%%%%%%%%%%%
\section{Model and dynamics}
\begin{figure}[h]
		\centering
		\includegraphics[scale=0.6]{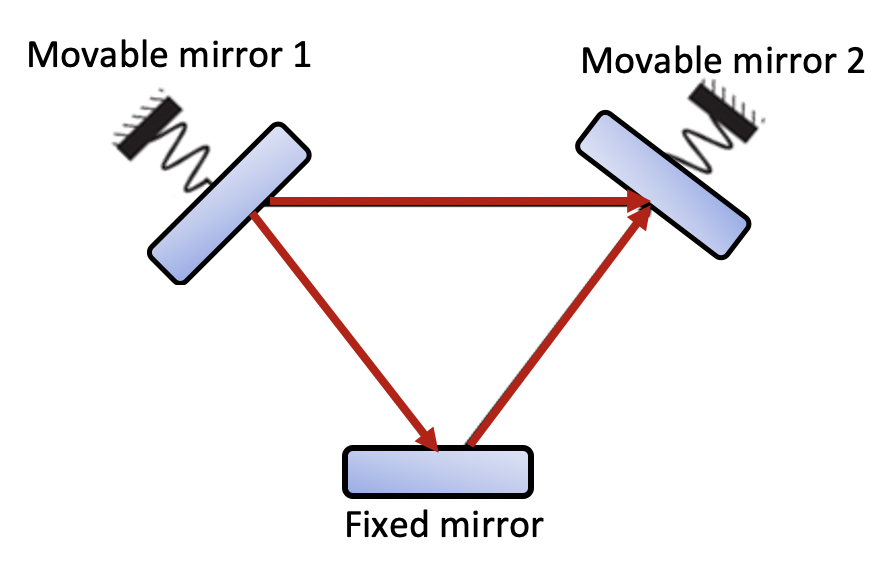}
		\caption{Schematic illustration of an optomechanical ring cavity driven by a coherent laser at a specific frequency $\Omega_{L}$.} 
		\label{ring}
	\end{figure}

%%%%%%%%%%%%%%%%%%%%%%%%%%%%%%%%%%%%%%%%%%%%%%%%%%%%%%%%%%%%%%%%%%%%%%%%%%%%
The system considered here is an optomechanical ring cavity with an arm length $l$ and frequency $\Omega_{c}$, as shown in Fig.(\ref{ring}). The cavity contains three mirrors in a triangular design: the fixed mirror partially transmits light, and the two movable ones are perfectly reflecting. These latter can be considered as quantum harmonic oscillators with effective masses $M_1$, $M_2$, and frequencies $\Omega_{m1}$ and $\Omega_{m2}$, respectively.

In this hybrid optomechanical system, the motion of the two mechanical oscillators is coupled to the cavity field via the radiation pressure force \cite{caves1980quantum,corbitt2006measurement}. This force causes a small displacement of the movable mirrors, enabling a quantum correlation between the optical mode and the two mechanical oscillators. In this background, the total Hamiltonian of the coupled system can be expressed as: $H = H_0 + H_{\text{int}} + H_{\text{drive}}$, which will be discussed in detail in the next section.
%%%%%%%%%%%%%%%%%%%%%%%%%%%%%%%%%%%%%%%%%%%%%%%%%%%%%%%%%%%%%%%%%%%%%%%%%%%
	\subsection{Hamiltonian}
	The system's Hamiltonian is defined as 
 
	\begin{equation}
	\begin{aligned}
	H/\hbar & = \Delta_{c} a^{\dagger}a + \frac{ \Omega_{m1}}{2}(q_{1}^{2} + p_{1}^{2}) + \frac{ \Omega_{m2}}{2}(q_{2}^{2} + p_{2}^{2}) + \lambda q_{1} q_{2}  + (\eta_{1} q_{1} - \eta_{2} q_{2}) a^{\dagger} a + i (a^{\dagger} - a) \varepsilon_L, 
	\end{aligned}
	\label{Hamil}
	\end{equation} 
where $\Delta_{c} = \Omega_{c} - \Omega_{L}$ represents the detuning between the cavity and the laser radiation, $a$ and $a^{\dagger}$ are, respectively, the annihilation and creation operators of the optical mode. Moreover, $g_i = \frac{\Omega_{c}}{l} \sqrt{\frac{\hbar}{M_i \Omega_{Mi}}}$ is the optomechanical coefficient given in units of s$^{-1}$. The parameter $\varepsilon_L$ describes the laser driving input by $\sqrt{\frac{2 \kappa_p P_L}{\hbar \Omega_{L}}}$, where $\kappa_p$ is the photon decay rate due to leakage from the cavity. The energy of the mechanical resonator is described in Eq.(\ref{Hamil}) by the second and third terms. We define $q_{i}$ and $p_{i}$ ($i = 1,2$) as, respectively, the dimensionless position and momentum operators of the $i$th mechanical quantum harmonic oscillator with $[q_{i}, p_{j}] = i \delta_{ij}$ ($i,j = 1, 2$). In addition, the interaction between the two movable mirrors is taken into consideration by the term $\lambda q_{1} q_{2}$ where $\lambda$ represents the coupling strength. We point out the interaction between the two movable mirrors and the optical mode caused by the radiation pressure, with $\eta_{i} = g_i \cos^{2}(\theta /2)$ ($i = 1,2$). It is worth mentioning that, the angle between the incident and the reflected light on the surfaces of the movable mirrors is denoted by $\theta$ (see Fig.(\ref{ring})).
	%%%%%%%%%%%%%%%%%%%%%%%%%%%%%%%%%%%%%%%%%%%%%%%%%%%%%%%%%%%%%%%%%%%%%%%%%%%%%%%%%%%%%%%%%%%%%%%%%
	\subsection{Quantum Langevin Equations}

A thorough analysis of the dynamics of the cavity field and the two mechanical modes must account for system dissipations and noise. This can be achieved by considering the Heisenberg-Langevin equations of motion (QLEs), which are derived from Eq.(\ref{Hamil}). By incorporating the fluctuation-dissipation terms, we obtain the following set of nonlinear equations:
	\begin{equation}
	\begin{aligned}
	\dot{q}_{1} &= \Omega_{M1} \ p_{1}, \\
	\dot{p}_{1} &= -\Omega_{M1} q_{1} - \gamma_{M1} \ p_{1} - \lambda q_{2} - \eta_{1} a^{\dagger}a + \xi_{1}, \\ 
	\dot{q}_{2} &= \Omega_{M2} \ p_{2}, \\
	\dot{p}_{2} &= -\omega_{M2} q_{2} - \gamma_{M2} \ p_{2} - \lambda q_{1} - \eta_{2} a^{\dagger} a + \xi_{2}, \\ 
	\dot{a} &= -(\kappa_p + i \Delta_{c})a - i(\eta_{1} q_{1} - \eta_{2} q_{2})a + \varepsilon_L +  \sqrt{2 \kappa_p}\, \nu_{in},\\
	\dot{a}^\dagger &= -(\kappa_p - i \Delta_{c}) a^{\dagger} + i(\eta_{1} q_{1} - \eta_{2} q_{2})a^{\dagger}  + \varepsilon_L +  \sqrt{2 \kappa_p}\, \nu_{in}^{\dagger},\\
	\end{aligned}
	\label{eq2}
	\end{equation}
	We have introduced the Brownian noise, denoted by $\xi_{i}(i = 1,2)$, which represents the interaction between the movable mirror and its environment. This noise has a zero mean and is defined by the following correlation function \cite{benguria1981quantum}:  
	\begin{equation}
	\begin{aligned}
	\langle \xi_{i}(t) \xi_{i} (t^{'}) + \xi_{i} (t^{'}) \xi_{i}(t) \rangle / 2 \simeq \gamma_{m_{i}}(2 n_{th_{i}} + 1) \delta(t - t^{'}),
	\end{aligned}
	\end{equation}
	with $\gamma_{M_{i}}$ is the damping rate representing the loss of mechanical energy of the $i$th mirror (i = 1,2). $n_{th_{i}}$ is the $i$th thermal phonon
	number and $k_B$ is the Boltzmann constant. We have introduced the vacuum noise at the input  $\nu_{in}$ at temperature $T$, whose only nonzero correlation function is \cite{walls2012quantum}:
	\begin{equation}
	\begin{aligned}
	\langle \nu_{in}(t) \ \nu_{in}^\dagger(t^{'})\rangle = \delta(t-t^{'}).
	\end{aligned}
	\end{equation}
	Equations \ref{eq2} are in general not solvable analytically. To simplify this issue, we use the scheme given in Ref. \cite{walls2008quantum}. A clear illustration is given in the next section. 
 
%%%%%%%%%%%%%%%%%%%%%%%%%%%%%%%%%%%%%%%%%%%%%%%%%%%%%%%%%%%%%%%%%%%%%
	\subsection{Linearization of QLEs}
	To analyze the dynamics of the coupled system, we apply a linearization approach by expressing each Heisenberg operator as the sum of its classical steady-state value and a zero-mean fluctuation operator \cite{vitali2007optomechanical}:
	\begin{equation}
	\begin{aligned}
	a &= a^s + \delta a, \\
	q_1 &= q_1^s + \delta q_1,\\
	p_1 &= p_1^s + \delta p_1,\\
	q_2 &= q_2^s + \delta q_2,\\
	p_2 &= p_2^s + \delta p_2.\\
	\end{aligned}
	\label{5}
	\end{equation}
	The corresponding steady-state values are presented by 
	\begin{equation}
	 p_1^s = 0,\hspace{0.5cm}
	 p_2^s = 0,\hspace{0.5cm} q_1^s = \frac{- \eta_1 |a^s|^2}{\Omega_{M_{1}}},\hspace{0.5cm} q_2^s = \frac{\eta_2 |a^s|^2}{\Omega_{M_{2}}}, \hspace{0.5cm} a^s = \frac{\varepsilon_L}{\kappa_p + i \Delta}, 
	\end{equation}
	where $\Delta = \Delta_{c} + (\eta_1 q_{1}^s - \eta_2 q_{2}^s)$. 
	
	Inserting Eqs. (\ref{5}) in Eqs. (\ref{eq2}), and introducing $ \delta x = \frac{\delta a + \delta a^{\dagger}}{\sqrt{2}}$ ,\; $\delta y = \frac{\delta a - \delta a^{\dagger}}{i \sqrt{2}}$ , \; $\delta x_{in} = \frac{\delta \nu_{in} + \delta \nu_{in}^{\dagger}}{\sqrt{2}}$ , \; $\delta y_{in} = \frac{\delta \nu_{in} - \delta \nu_{in}^{\dagger}}{i \sqrt{2}}$
	allows to obtain the following linearized Langevin equations:
	\begin{equation}
	\begin{aligned}
	\delta \dot{q}_{1} &= \Omega_{M_{1}} \delta p_{1}, \\
	\delta \dot{p}_{1} &= - \Omega_{M_{1}} \delta q_{1}  - \gamma_{M1} \ \delta p_{1} - \lambda \delta q_{2} - G_1 \delta x+ \xi_{1}, \\
	\delta \dot{q}_{2} &= \Omega_{M_{2}} \ \delta p_{2}, \\
	\delta \dot{p}_{2} &= - \Omega_{M_{2}} \delta q_{2}  - \gamma_{M2} \ \delta p_{2} - \lambda \delta q_{1} + G_2 \delta x  + \xi_{2},\\
	\delta \dot{x} &= - \kappa_p \delta x + \Delta \delta y + \sqrt{2 \kappa_p}\, \delta x_{in}, \\
	\delta \dot{y} &= G_1 \delta q_{1} +  G_2 \delta q_{2} - \kappa_p \delta y - \Delta \delta x + \sqrt{2 \kappa_p}\, \delta x_{in}, \\
	\end{aligned}
	\label{eq7}
	\end{equation}
	with $G_1 =  \sqrt{2} \eta_1 a^s $ and $G_2 =  \sqrt{2} \eta_2 a^s $.
	The resulting system of equations in (\ref{eq7}) can be represented in the matrix form: 
	\begin{equation}
	\begin{aligned}
	\dot{v}(t) = S v(t) + F(t),
	\end{aligned}
	\label{eq8}
	\end{equation}
	where $v(t)$ and $F(t)$ represent the column vector of fluctuations and the column vector of noise operators, respectively, with their transposes given by:
	
	\begin{eqnarray}
	v^{T}(\infty) &=& (\delta q_{1}(\infty),\, \delta p_{1}(\infty),\, \delta q_{2}(\infty),\, \delta p_{2}(\infty), \, \delta x(\infty), \, \delta y(\infty)) \nonumber \\ 
	F^T(t) &=& (0, \xi_{1},0,  \xi_{2}, \sqrt{2 \kappa}\, \delta x_{in}, \sqrt{2 \kappa}\, \delta y_{in}).
	\end{eqnarray}
	The drift matrix $S$, gets
	\begin{equation}
	S = \begin{pmatrix} 0 &  \Omega_{M_{1}} & 	0 & 0 & 0 & 0 \\ - \Omega_{m_{1}} &  - \gamma_{M_{1}} & - \lambda & 0 & -G_{1} & 0 \\  0 & 0 & 0 & \Omega_{M_{2}} & 0 & 0 \\ - \lambda & 0 & -\Omega_{M_{2}} &  - \gamma_{M_{2}} & G_{2} & 0 \\  0 & 0 & 0 & 0 & -\kappa_p & \Delta \\
	-G_{1} & 0 & G_{2} & 0 & -\Delta & -\kappa_p \\
	\end{pmatrix}.
	\end{equation}
	The solution of the differential equation (\ref{eq8}), is $u(t) = Y(t)u(0) + \int_{0}^{t} dx Y(x) \eta(t - x)  $, with $Y(t) = \exp{(At)}$. 
	%%%%%%%%%%%%%%%%%%%%%%%%%%%%%%%%%%%%%%%%%%%%%%%%%%%%%%%%%%%%%%%%%%%%%%%%%%%%%%%%%%%%
	\subsection{Covariance matrix}
	When the Routh–Hurwitz criteria are met for stability \cite{vitali2007optomechanical, dejesus1987routh, mari2009gently}, the steady-state covariance matrix $Cm$ fulfills the following Lyapunov equation: 

	\begin{equation}
	\begin{aligned}
	S \, Cm + Cm \, S^{T} =  -N,
	\end{aligned}
	\label{eq9}
	\end{equation}
	where $N$ represents the diagonal matrix of noise correlations; it is given by
	\begin{equation}
	\begin{aligned}
	N_{ij} \delta (t - t^{'}) = \frac{1}{2} (\left\langle F_i(t) F_j(t^{'}) + F_j(t^{'}) F_i(t)\right\rangle),
	\end{aligned}
	\end{equation}
	Therefore, we get:
	\begin{equation}
	\begin{aligned}
	N = \begin{pmatrix} 0 & 0 & 0 & 0 & 0 & 0 \\  
	0 & \gamma_{M_{1}}(2 n_{th_{1}} + 1) & 0 & 0 & 0 & 0 \\
	0 & 0 & 0 & 0 & 0 & 0 \\  
	0 & 0 & 0 & \gamma_{M_{2}}(2 n_{th_{2}} + 1) & 0 & 0 \\ 
	0 & 0 & 0 & 0 &  \kappa_p & 0 \\  
	0 & 0 & 0 & 0 & 0 & \kappa_p   \\ 
	\end{pmatrix}
	\end{aligned}.
	\end{equation}
	The covariance matrix $Cm$ can be expressed in a block form:
	\begin{equation}
	\begin{aligned}
	Cm = \begin{pmatrix} C_{M1} & C_{M1M2} & C_{M1op} \\  
	C_{M1M2}^T & C_{M2} & C_{M2op}  \\ 
	C_{M1op}^T & C_{M2op}^T & C_{op} \\ 
	\end{pmatrix}
	\end{aligned}
	\label{matrv}
	\end{equation}
	where $C_{M_{j}} (j = 1,2)$ is the covariance matrix of the $j$th mechanical mode, and $C_{op}$ of the optical mode, respectively.
	
	This paper examines the behavior of tripartite and bipartite entanglement. When analyzing the behavior of only two subsystems, the global $6 \times 6$ covariance matrix $Cm$ of Eq. (\ref{matrv}) can be reduced to a $4 \times 4$ submatrix $C_S$, including only the covariance matrices of the subsystems of interest:
	\begin{equation}
	\begin{aligned}
	C_S = \begin{pmatrix} A_s &  K_s \\  
	K_s^T & B_s \\  
	\end{pmatrix}
	\end{aligned}
	\end{equation}
	$A_s$ and $B_s$ are the $2 \times 2$ covariance matrices representing the single modes, while $K_s$ is the $2 \times 2$ covariance matrix representing the quantum correlations between the two subsystems.
	%%%%%%%%%%%%%%%%%%%%%%%%%%%%%%%%%%%%%%%%%%%%%%%%%%%%%%%%%%%%%%%%%%%%%%%%%%%%%%
	\section{Entanglement analysis}
	Entanglement is at the heart of a variety of quantum information applications such as quantum metrology, quantum cryptography, and quantum communication \cite{breuer2002theory,horodecki2009quantum,zou2021quantum}. Recently, the control of entanglement dynamics in a tripartite coupled system has been investigated \cite{gonzalez2017control}. Here, we study the stationary entanglement presented in a coupled optomechanical system between photon and phonon modes. 

%%%%%%%%%%%%%%%%%%%%%%%%%%%%%%%%%%%%%%%%%%%%%%%%%%%%%%%%%%%%%%%%%%%%%%%%%%%%%%%%%%%%%%%%%%%%%%%%%%%%%%%%%%%%%%%%
	\subsection{Bipartite entanglement}
%%%%%%%%%%%%%%%%%%%%%%%%%%%%%%%%%%%%%%%%%%%%%%%%%%%%%%%%%%%%%%%%%%%%%%%%%%%%%%%%%%%%%%%%%%%%%%%%%%%%%%%%%%%%%%%
	We numerically investigate the stationary entanglement between the optical mode and each mechanical mode, as well as between the two mechanical modes. To quantify this entanglement, we use the logarithmic negativity $E_N$, a widely recognized measure of quantum entanglement, defined for Gaussian continuous variable systems as follows \cite{vidal2002computable,plenio2005logarithmic}:
	\begin{equation}
	\begin{aligned}
	E_{N} = \max [0, -\ln{ 2\tilde{\mu}_-}] ,
	\end{aligned}
	\end{equation}
	with 
	\begin{equation}
	\begin{aligned}
	\tilde{\mu}_- = \min\left\{  \hbox{eig} \ | [\oplus_{j=1}^2 (-\sigma_y)] P C_{S} P |\right\},
	\end{aligned}
	\end{equation}
	where $\sigma_y$ is the y-Pauli matrix, $C_{S}$ is the $4 \times 4$ covariance matrix of the two subsystems and $P = \sigma_z \oplus 1$, with $\sigma_z$ being the z-Pauli matrix. The two modes are entangled when $E_N >0$, which is equivalent to $\tilde{\nu}_- < \frac{1}{2}$. If not, the states are separable, which is compatible with Simon's necessary and sufficient criterion for bipartite entanglement \cite{simon2000peres}. 
	
	In order to evaluate the logarithmic negativity, some parameters are taken from the most pertinent experiments \cite{groblacher2009observation}, where the angle $\theta = \frac{\pi}{3}$, the laser frequency $\Omega_{L} = 2 \pi \times \, 3.7 \times 10^{14}$ Hz, the frequencies $\Omega_{M1} = \Omega_{M2} = \Omega_{M} = 2 \pi \times 10^7$ Hz, $\gamma_{M1} = \gamma_{M2} = \gamma_{M} = 2 \pi \times 10^2$ Hz, $g_1 = 2 \pi \, 35$Hz, $g_2 = 0.9 g_1$ and $\kappa_p = \pi \times 10^7 Hz$.
	
	\begin{figure}[h]
		\centering
		\includegraphics[scale=0.55]{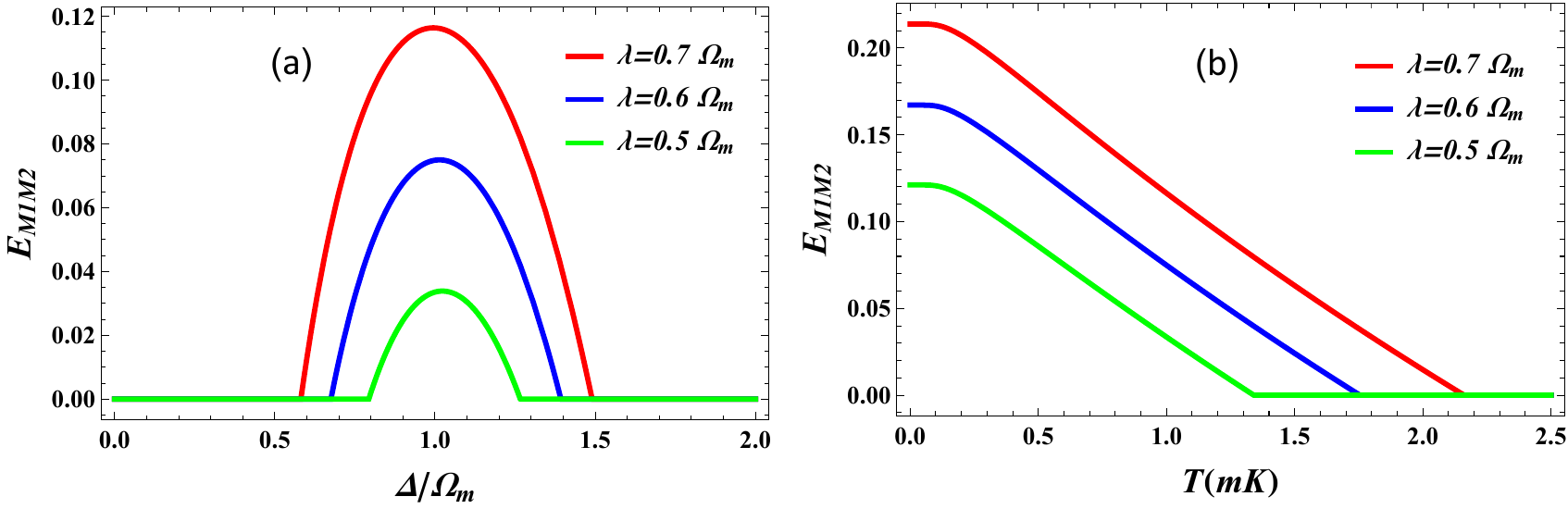}
		\caption{(a) $E_{M1M2}$ versus $\Delta/\Omega_M$ for different values of $\lambda$. (b) $E_{M1M2}$ versus $T$ for different values of $\lambda$.} 
		\label{fig1}
	\end{figure}
	
	Figure \ref{fig1}(a) shows the entanglement between the two mechanical modes, i.e. $E_{M1M2}$ as a function of the normalized detuning $\Delta/\Omega_M$ for different values of the coupling $\lambda$. Interestingly, the quantum entanglement exists only within a specific finite range of values for $\Delta$ around $\omega_M$, where mechanical entanglement reachs its maximum. This interval is a rich area for developing quantum information processing systems \cite{aspelmeyer2014cavity}. By understanding and controlling detuning, one can manipulate entanglement properties. We see also from the figure that increasing the coupling strength $\lambda$ not only strengthens the entanglement but also broadens the range of effective detuning where entanglement exists. This is highly significant because stronger entanglement and a broader effective detuning range make the phenomenon easier to achieve and observe in experiments \cite{palomaki2013entangling,vitali2007optomechanical,riedinger2018remote}. The fact that higher coupling strength $\lambda$ leads to stronger and wider width of entanglement region is due to the fact that, the two mechanical resonators are directly coupled, which is typically modeled in the Hamiltonian that describes direct interactions between the dimensionless position operators of the mechanical resonators. 
	
In figure \ref{fig1}(b), we plot the entanglement between the two mechanical mirrors as a function of the temperature for different values of the coupling $\lambda$. The level of entanglement diminishes with increasing temperature, as expected, due to fundamental reasons related to thermal noise and decoherence. Indeed, when the thermal bath temperature increases, the mechanical resonator's thermal phonon number increases, which introduces noise and thermal fluctuations in the system. Subsequently, the thermal fluctuations interfere with quantum correlations allowing for entanglement. It is important to note that with the increase of $\lambda$, entanglement persists up to a much higher value. Adjusting and optimizing the coupling between the two mirrors might contribute to a robust entanglement, resulting in variety of applications in quantum information processing and communication \cite{aspelmeyer2014cavity,meystre2017cavity}. 

	\begin{figure}[h]
		\centering
		\includegraphics[scale=0.55]{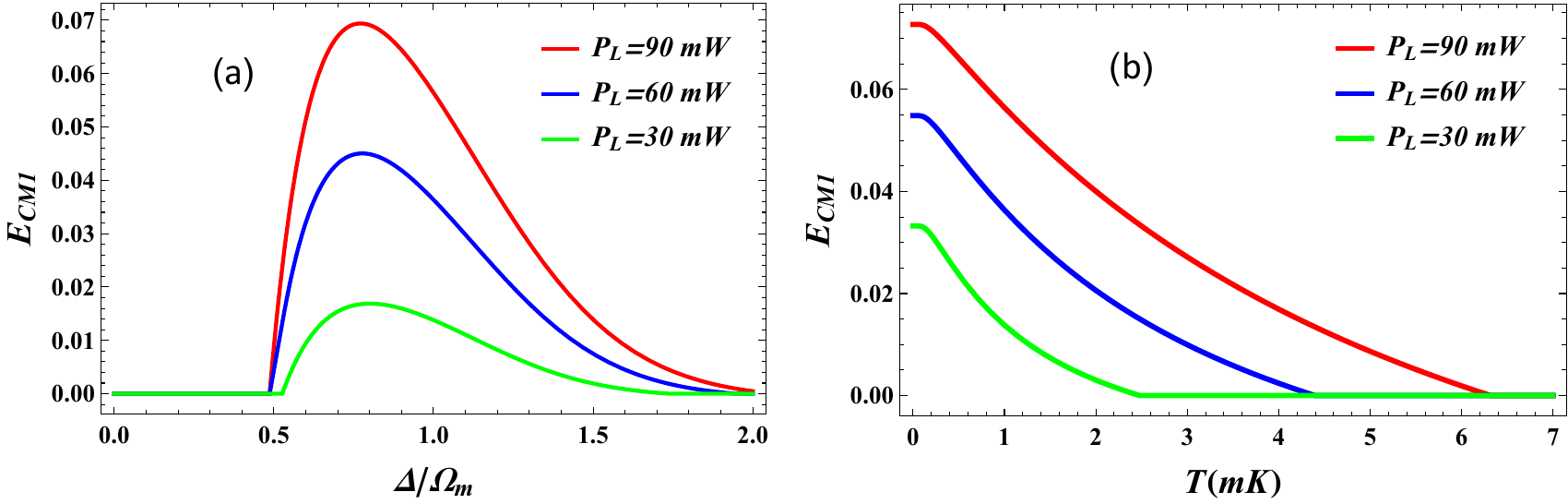}
		\caption{(a) $E_{CM1}$ versus $\Delta/\Omega_M$ for different values of $P_L$. (b) $E_{CM1}$ versus $T$ for different values of $P_L$.} 
		\label{fig2}
	\end{figure}

Next, to investigate how entanglement depends on various other parameters such as the laser power $P_L$, we show in Figure \ref{fig2} the entanglement between the movable mirror 1 and the optical mode. Figure \ref{fig2}(a) shows the logarithmic negativity $E_{CM1}$ versus the normalized detuning $\Delta/\Omega_M$. Quantum entanglement between the mechanical and the optical mode $1$ exists only within a limited range of values of $\Delta$ around $\Delta \sim 0.8 \Omega_M$. In addition, it is remarkable that, the largest region of entanglement corresponds to the chosen parameter of power $P_L = 90$ mW, which means that the stationary optomechanical entanglement increases as the laser power increases. Strong entanglement can be achieved with significantly higher driving power. However, increasing the power $P_L$ is limited by stability conditions, since uncontrolled increases can cause instabilities within the system. Denotes that, when the cavity is coherently driven by a strong external laser field characterized by a great number of photons. The latter apply a radiation pressure force on the surfaces of the movable mirrors, proportional to the instantaneous photon number within the cavity. This, in turn, amplifies the intensity of the radiation pressure force itself and allows to a strong photon-phonon coupling and therefore strong stationary optomechanical entanglement.

	\begin{figure}[h]
		\centering
		\includegraphics[scale=0.55]{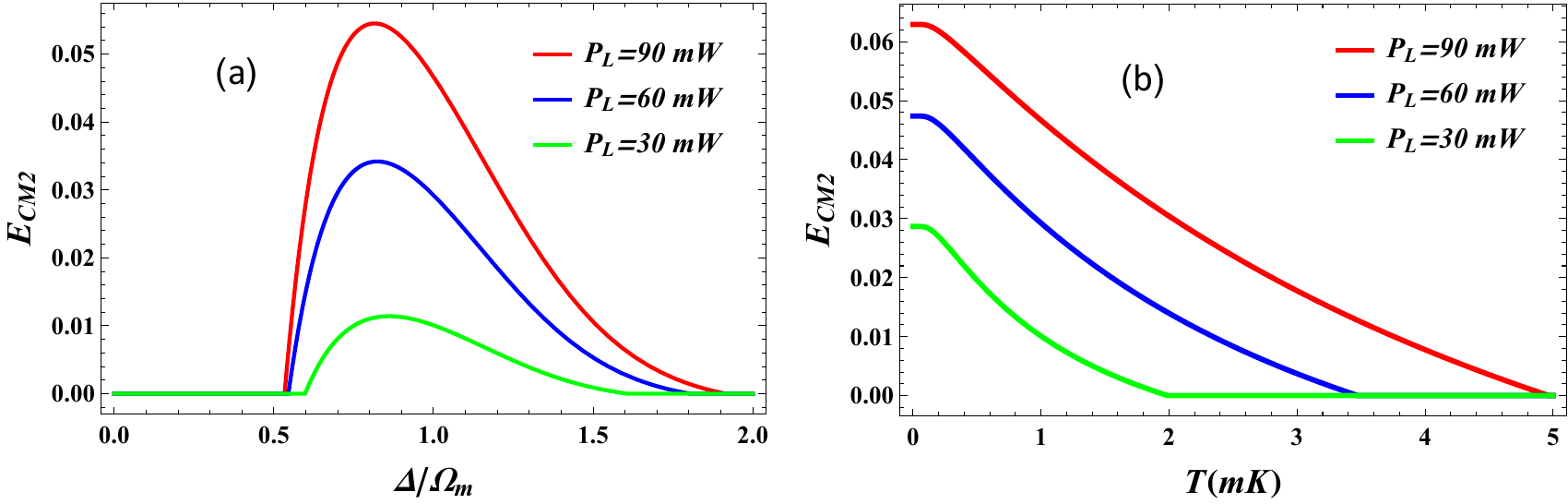}
		\caption{(a) $E_{CM2}$ versus $\Delta/\Omega_M$ for different values of $P_L$. (b) $E_{CM2}$ versus $T$ for different values of $P_L$. } 
		\label{fig3}
	\end{figure}

In figure \ref{fig2}(b), we plot the optomechanical (cavity-mirror 1) stationary entanglement as a function of the temperature. For a fixed temperature, e.g., for $T = 1$ mK and when $P_L = 90$ mW, the numerical value of the entanglement is $E_{CM1} \simeq 0.056$, whereas, for $P_L = 60$ mW, it is $E_{CM1} \simeq 0.035$. Moreover, for $P_L = 30$ mW, the numerical value of the entanglement is $E_{CM1} \simeq 0.013$. This illustrates the impact of the driving laser power on entanglement, influenced by its relationship with optomechanical coupling. Accordingly, The laser power positively influences the quantum correlations, as measured by the logarithmic negativity. It is worth mentioning that, for higher values of laser power $P_L$, entanglement persists up to a much higher value of temperature. This is due to the robustness against thermal bath environment-induced decoherence, since the movable mirror is subject to thermal fluctuations from its bath. In fact, each mirror is subject to quantum Brownian noise due to its coupling to its own environment. In addition, another source of fluctuations arises from the flapping of the pump photons with the thermal phonons.       
	
For a comprehensive review of the system, we analyze the impact of the cavity field on the second mirror. In this background, we plot in Figure \ref{fig3} the logarithmic negativity $E_{CM2}$ as a function of the normalized detuning $\Delta/\Omega_M$ and the bath temperature $T(mK)$ for a range of values of the laser power $P_L$. As expected, we can see similar behavior with $E_{CM1}$, for instance, as $T$ increases $E_{CM2}$ decrease until reaching zero. This explain that, a higher thermal energy means that the mechanical mode is more prone to interact with its environment, leading to an increase of the decoherence. This later is the technics through which a quantum system dissipates its quantum characteristics and starts to behave in a more classical manner \cite{verhagen2012quantum}. Comparing Figures \ref{fig2} and  \ref{fig3}, we remark that, the entanglement between the optical mode and mirror 1 is higher that the entanglement between the optical mode and the second mirror. For example when $P_L = 90$ mW and $T = 1$ mK, we have $E_{CM1} = 0.055$ whereas $E_{CM2} = 0.048$. This difference is mainly due to the fact that $g_1>g_2$. The asymetry between the two optomechanical coupling can be related to many reasons, such as the difference of mass between the two mirrors or the geometry of the ring cavity. These factors among others can cause the entanglement between each mirror and the optical mode to be different. 
%%%%%%%%%%%%%%%%%%%%%%%%%%%%%%%%%%%%%%%%%%%%%%%%%%%%%%%%%%%%%%%%%%%%%%%%%%%%%%%%%%%%%%%%%%%%%
	\subsection{Tripartite entanglement}
	Besides bipartite entanglement, the tripartite entanglement among the three modes of the coupled optomechanical system can also be investigated. In fact, numerous quantifiers of genuine tripatite entanglement in multimode Gaussian states have been proposed, for example, the tripartite negativity \cite{zhang2015more} and minimal residual contangle \cite{adesso2006continuous}. Here, we aim at shedding on the minimum residual contangle, as a measure of tripartite entanglement, which is a continuous variable analogue of the tangle for discrete-variable systems \cite{li2018magnon}. Therefore, the bona fide quantification of tripartite entanglement is given by
	\begin{equation}
	\begin{aligned}
	R_{min} = \min {R_{1|23}, \ R_{2|31}, \ R_{3|12}},
	\end{aligned}
	\end{equation}
where $R_{A|BC} = C_{A|BC} - C_{A|B} - C_{A|C} (A,B,C = 1,2,3$ with $B=A+1$ for $A=1,2, B=1$ for $A=3$ and $C \neq A,B$) is the residual contangle, and $C_{x|y}$ is the contangle of subsystems of $x$ and $y$, which is a proper entanglement defined as the squared logarithmic negativity:
	
	\begin{equation}
	\begin{aligned}
	C_{x|y} = (E_N^{x|y})^2,
	\end{aligned}
	\end{equation}
with  $E_N^{A|BC} = \max [0 , -\ln 2 \mu_{A|BC}]$ is the logarithmic negativity of the one mode-versus-two modes bipartitions in the system, where $\mu_{A|BC} = \min\left\{  \hbox{eig} \ | [\oplus_{j=1}^3 (-\sigma_y)] P_{A|BC} \ V \ P_{A|BC} |\right\}$, with $P_{A|BC} = \sigma_z \oplus 1 \oplus 1$, $P_{B|AC} = 1 \oplus \sigma_z \oplus  1$ and $P_{C|AB} = 1 \oplus 1 \oplus \sigma_z $ are the matrices of the partial transposition of the tripartite covariance matrix, and $V$ is the $6 \times 6$ covariance matrix of the coupled optomechanical system (Eq. \ref{matrv}).
	
	\begin{figure}[h]
		\centering
		\includegraphics[scale=0.75]{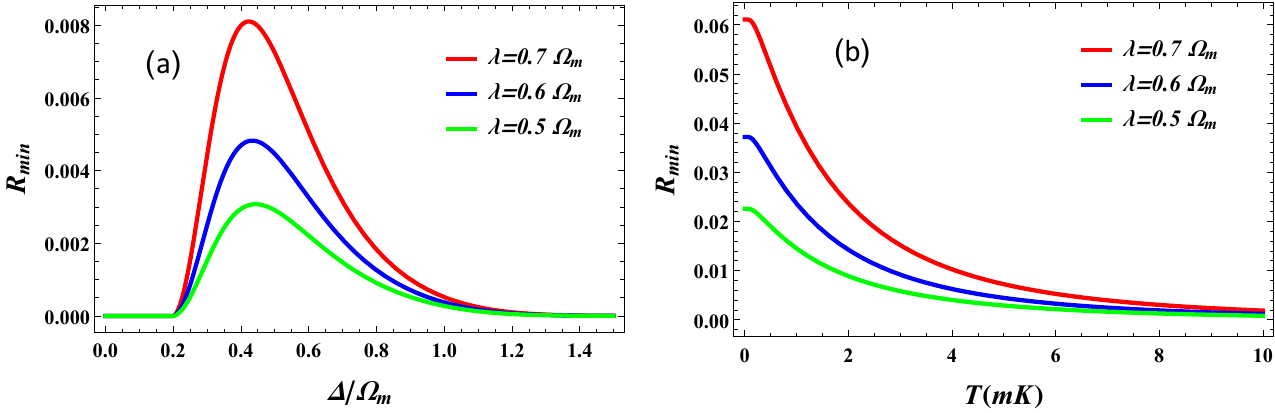}
		\caption{Residual contangle $R_{min}$ as a function of the normalized detuning $\Delta/\Omega_M$ and the bath temperature $T$ (b) across different values of $\lambda$.} 
		\label{fig4}
	\end{figure}
	
For a clear illustration of the behavior of the tripartite entanglement between the two mechanical modes and the optical mode, we plot in Figure \ref{fig4} the residual contangle $R_{min}$ as a function of the normalized detuning $\Delta/\Omega_M$ and the bath temperature $T$ for different values of $\lambda$. In Figure \ref{fig4}(a), we depict the contangle as a function of $\Delta/\Omega_M$. Noteworthy, the largest and highest region is the one corresponding to the greatest value of $\lambda$. In other words, $R_{min}$ increases as the coupling $\lambda$ increases. Enhancing the coupling provide an increase of tripartite entanglement. In addition, using a ring configuration offers several specific advantages and allows to have a robust entanglement due to the interconnected systems that exchange energy in a cyclic manner. Indeed, the ring configuration strengthens the optomechanical coupling through the motion of the two movable mirrors. Entanglement arises from the influence of the mirror dynamics, because when considering a fixed mirror, entanglement would not occur. Besides, ring cavities can confine light effectively and support strong interactions between optical and mechanical modes that make it a preferred choice in generating stationary correlations.
	
\begin{figure}[h]
		\centering
		\includegraphics[scale=0.75]{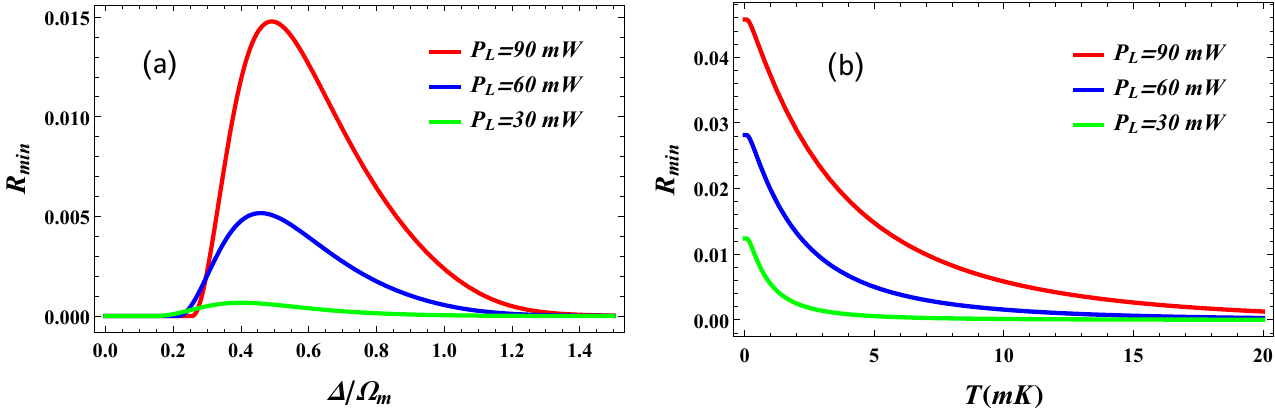}
		\caption{Residual contangle $R_{min}$ as a function of the normalized detuning $\Delta/\Omega_M$ and the bath temperature $T$ (b) at distinct values of the pumping power $P_L$.} 
		\label{fig5}
	\end{figure}

In Figure \ref{fig4}(b), the system tripartite entanglement, as quantified by $R_{min}$ decreases as a function of temperature. This later is influenced by thermal excitations from quantum mechanical effects. In fact, the residual cotangle decrease because thermal noise introduces decoherence, disrupting quantum correlations. At high temperatures, the residual cotangle values approach zero, signifying the full destruction of quantum entanglement by thermal noise. For a comprehensive view of the residual contangle, we plot, in Figure \ref{fig5}, $R_{min}$ as a function of the normalized detuning $\Delta/\Omega_M$ and the bath temperature $T$ under varying values of the pumping power $P_L$.    
	
	Figure \ref{fig5}(a) shows the residual contangle $R_{min}$ against the normalized detuning $\Delta/\Omega_M$ for three different pumping powers: $P_L = 30$ mW, $P_L = 60$ mW, and $P_L = 90$ mW. For all three pumping powers, $R_{min}$ decreases for higher values of detuning as $\Delta/\Omega_M$ increases, indicating that entanglement decreases with increasing detuning. At very low detuning values, higher pumping power $(P_L= 90 mW)$ results in higher $R_{min}$, indicating stronger tripartite entanglement. In this background, we conclude that, the convenient choice of detuning for having optimal tripartite entanglement corresponds to $\Delta \simeq 0.5\Omega_M$ and higher pumping powers. Higher pumping power is required to have a nonzero residual contangle. Furthermore, the pumping power has a favorable effect on entanglement since it enlarges the interval of $\Delta/\Omega_M$ where entanglement exist.
	
In figure \ref{fig5}(b), we plot the residual cotangle $R_{min}$ against temperature $T(mK)$ for the same three pumping powers. $R_{min}$ decreases with increasing temperature for all pumping powers, reflecting the detrimental effect of thermal noise on entanglement. At lower temperatures, higher pumping power results in higher $R_{min}$, suggesting that stronger pumping power grant a robust entanglement against thermal fluctuations. As temperature increases, $R_{min}$ for all pumping powers goes to zero, indicating that thermal noise eventually dominates over the effects of pumping power.
	
	In summary, the balance between detuning, couplings, thermal effects, and pumping power is required in maintaining entanglement. These insights are crucial for designing and optimizing optomechanical systems for quantum technologies.

	\section{Conclusion}
	
In this work, we have delved into the dynamics of bipartite and tripartite entanglement within a ring optomechanical cavity, shedding light on the stationary quantum correlations between two mechanical modes and optical mode. The system consists of a ring cavity composed of a fixed mirror and two movable ones, pumped with a coherent laser source to enhance the generated quantum entanglement. We conducted a thorough analysis of the coupled system's dynamics, which allowed us to linearize the set of quantum Langevin equations. This linearization facilitated the derivation of the $6 \times 6$ steady-state covariance matrix, fully characterizing the ring optomechanical system. Afterwards, we presented an analysis of the bipartite entanglement shared by the coupled subsystems, measuring the amount of entanglement using logarithmic negativity. Through detailed numerical simulations, we demonstrated that bipartite entanglement between different pairs of modes can be robustly achieved and controlled by adjusting system parameters such as cavity detuning, mechanical coupling strength, and pumping power. For instance, we find that mechanical entanglement is optimized within specific detuning intervals and is robust against thermal noise at higher mechanical coupling strengths. Similarly, optomechanical entanglement is enhanced with increased laser power but remains vulnerable to temperature-induced decoherence. Furthermore, we quantified tripartite entanglement using the residual contangle, revealing conditions under which three-mode entangled states can be sustained. In fact, higher mechanical coupling strengths and laser powers generally lead to stronger tripartite entanglement, while thermal noise poses a significant challenge, particularly at elevated temperatures. The ability to manipulate tripartite entanglement opens new avenues for implementing more complex quantum protocols, including those requiring multipartite correlations. Our results also demonstrate that the coupling between optical and mechanical components in ring cavities can be finely tuned to generate significant bipartite and tripartite entanglement, an essential resource for quantum
applications, quantum information processing and quantum communication technologies. Finally, the study of bipartite and tripartite entanglement in ring optomechanical cavities not only enhances our understanding of fundamental quantum phenomena but also offers potential for practical applications in the rapidly evolving field of quantum technologies. Future work may focus on the investigation of the effects of other parameters on enhancement of entanglement.

\end{document}